\documentclass[aps,pre,twocolumn,superscriptaddress,showkeys,showpacs]{revtex4-1}

\newcommand{\checkVIB}[1]{\delta #1}
\newcommand{\widetildeFV}[1]{\delta #1}
\renewcommand{\d}{\mathrm{d}}
\newcommand{\lin}{l_{\text{in}}}
\newcommand{\leqq}{l_{\textup{eq}}}
\newcommand{\lbar}{\bar{l}}
\usepackage{graphicx}
\usepackage{amsmath,bm}
\usepackage{color}

\usepackage{fancyhdr}            
\fancypagestyle{plain}{%
  \fancyhead{}                                     
  \fancyfoot[CE,CO]{\thepage}       
  \fancyhead[CE,CO]{\textcolor[rgb]{0.64,0.15,0.15}{Published in \emph{Physical Review E} \textbf{94}, 063005
\\
doi: https://dx.doi.org/10.1103/PhysRevE.94.063005}}
\setlength{\headheight}{14.5pt}}

\def\salto#1#2{
[\mbox{\hspace{-#1em}}[#2]\mbox{\hspace{-#1em}}]}

\graphicspath{{figures/}}

\begin{document}


\title{Asymptotic self-restabilization of a continuous elastic structure}


\author{F. Bosi} \affiliation{DICAM - University of Trento, via Mesiano 77, I-38123 Trento, Italy}

\author{D. Misseroni} \affiliation{DICAM - University of Trento, via Mesiano 77, I-38123 Trento, Italy}
\author{F. Dal Corso} \affiliation{DICAM - University of Trento, via Mesiano 77, I-38123 Trento, Italy}
\author{S. Neukirch} \affiliation{Sorbonne Universit\'es, UPMC Univ Paris 06, CNRS, UMR 7190, Institut Jean Le Rond d'Alembert, F-75005 Paris, France}
\author{D. Bigoni}
\altaffiliation[Corresponding author: ]{bigoni@unitn.it; +39 0461 282507}
\affiliation{DICAM - University of Trento, via Mesiano 77, I-38123 Trento, Italy}


\date{\today}

\begin{abstract}
A challenge in soft robotics and soft actuation is the determination of an elastic system which spontaneously recovers its trivial path during postcritical deformation after a bifurcation. The interest in this behaviour is that a displacement component spontaneously cycles around a null value, thus
producing a cyclic soft mechanism.
An example of such a system is theoretically proven through the solution of the Elastica and a stability analysis based on dynamic perturbations. It is shown that the asymptotic self-restabilization
is driven by  the development of a configurational force, of similar nature to the Peach-Koehler interaction between dislocations in crystals, which is derived from the
principle of least action. A proof-of-concept prototype of the discovered elastic system is designed, realized, and tested, showing that this innovative behaviour can be obtained in a real mechanical apparatus.

\end{abstract}

%
%
%
%
%


\maketitle


%
%
%
%
%
%
%
\section{Introduction}
%
%
%
The straight configuration of an axially compressed elastic rod becomes unstable at buckling, so that a bent configuration emerges and the rod usually does not recover its straight configuration during the post-bifurcation deformation~\cite{Timoshenko-Gere-Libro:1961,Kuznetsov-Levyakov:2002}.
In the traditional mechanical design, the post-bifurcation behaviour, characterized by large deformations, is typically avoided to preserve mechanical integrity, but nowadays bio-inspired structures are often designed to
work in a large displacement regime \cite{nature,desimo,Laschi,Bigoni,esse2,Yang}. Therefore, an interesting challenge for applications to soft-robotics and compliant mechanisms
is to find a \lq self-restabilizing structure', in which the straight configuration is spontaneously recovered during loading after buckling.
Examples of these elastic systems have been provided, for which the instability region of the trivial path is bounded, becoming a sort of \lq island'. However, these restabilizations usually do not occur \lq spontaneously', so that the systems have to be \lq externally' moved back to the straight configuration (see the examples presented by Feodosyev~\cite{Feodosyev-Libro:1977} and by Bigoni et al.~\cite{Bigoni-Bosi-DalCorso-Misseroni-1:2014}, the former referred to a discrete system, the latter to a continuous elastic structure). For a two-degrees of freedom discrete system, spontaneous restabilization has been theoretically proven, but only in the presence of non-linear springs~\cite{PotierFerry:1987}.

The aim of the present article is to design, realize, and test a continuous elastic system which displays an \lq asymptotic self-restabilization' in the following sense: although bifurcation does not occur, because the system is imperfect, the deflection  initially grows  and subsequently decays up to vanish during a monotonically increasing loading.

\begin{figure}[tp]
  \begin{center}
      \includegraphics[width= 8.5 cm]{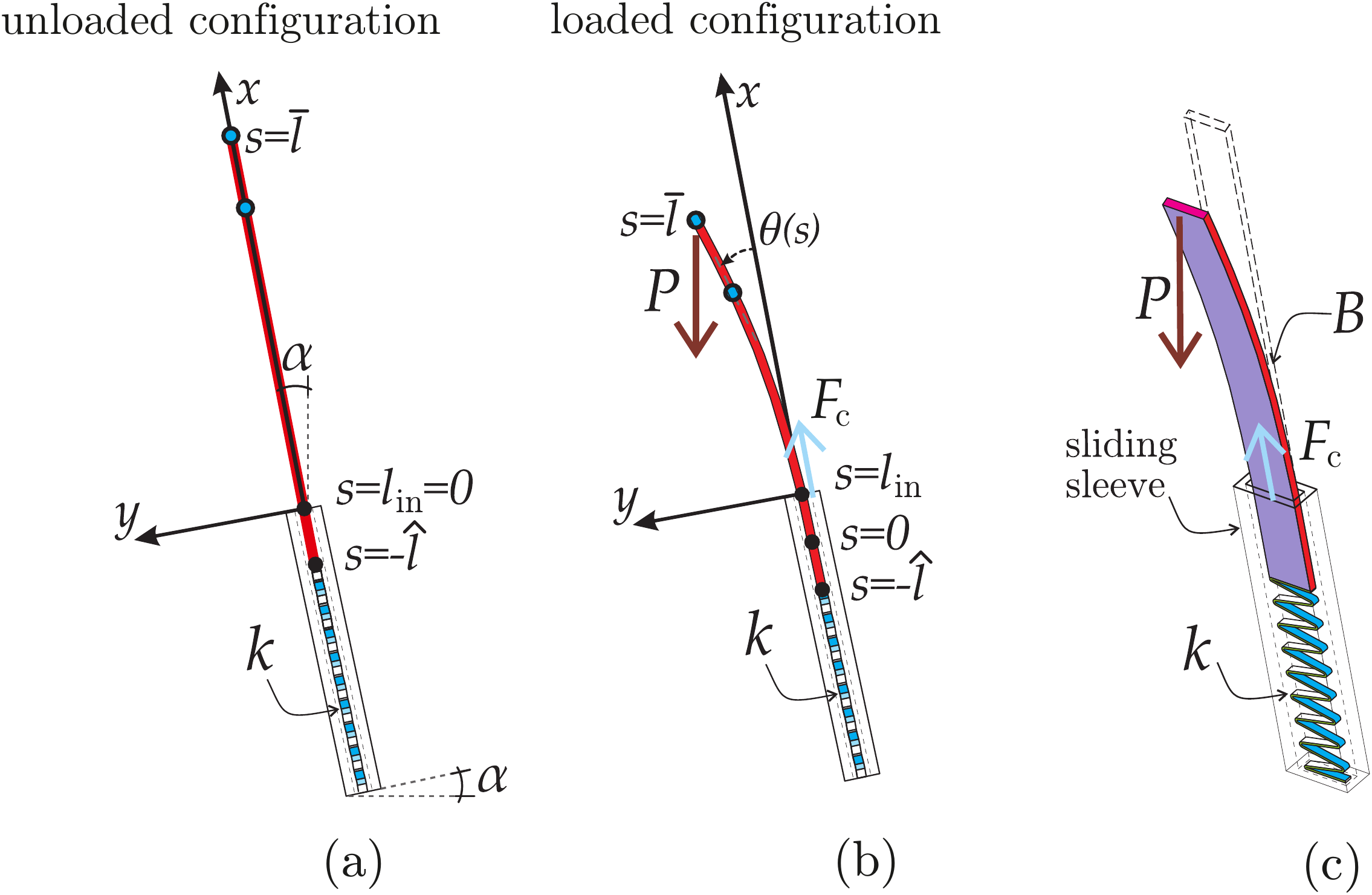}
\caption{\footnotesize An elastic rod of constant bending stiffness $B$ is loaded at its free end with a dead load $P$, while the other end can slide into a frictionless sleeve, against an axial linear spring of stiffness $k$.
The curvilinear coordinate $s$ measures the position along the rod, with $s=-\hat{l}$ corresponding to the rod's edge attached to the spring.
The total length of the rod is $\lbar+\hat{l}$, while the length of the rod inside the sliding sleeve is $\hat{l}+\lin$. The rotation of the rod's axis with respect to the the straight configuration is measured by $\theta(s)$. A configurational force $F_\textup{c}$ is developed at the sliding sleeve exit, $s=\lin$, whenever at this point the rod's curvature does not vanish.
The considered system is sketched in its undeformed configuration (A), in its deformed configuration (B), and in both configurations in a perspective view, with the undeformed configuration shown dashed (C). Due to the presence of the tilt angle $\alpha$, the system is an
imperfect version of the structure analyzed in ~\cite{Bigoni-Bosi-DalCorso-Misseroni-1:2014}. }
\label{systemrist}
  \end{center}
\end{figure}

The continuous elastic system is a planar, slender and inextensible rod, modeled as the elastica \cite{sebastiano, estrudimi,magnetico,basilio}. In particular, the developed system is an imperfect version of that analyzed in ~\cite{Bigoni-Bosi-DalCorso-Misseroni-1:2014}, where an elastic rod (with bending stiffness $B$) penetrates into a frictionless sliding sleeve and is restrained with a linear spring (of stiffness $k$). The imperfection is the tilt angle $\alpha$, which is null in the perfect case, see Fig.~\ref{systemrist}. A dead load $P$ is applied at the upper end of the elastic rod. The rod has a total length $\lbar+\hat{l}$,
of which the length $\hat{l}+\lin$ lies inside the sliding sleeve, with $\lin=0$ when $P=0$.

\begin{figure}[tp]
  \begin{center}
      \includegraphics[width= 8.5 cm]{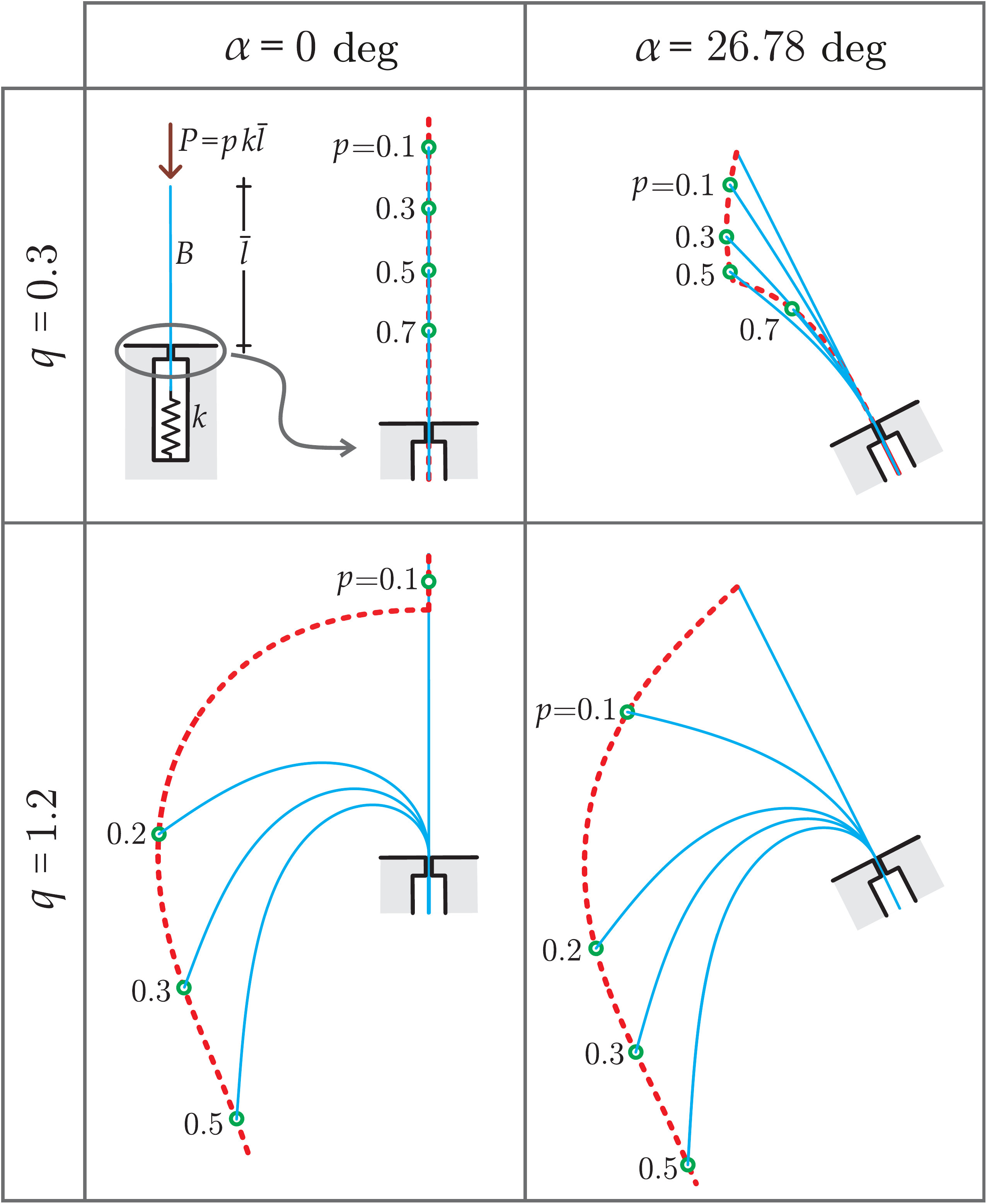}
\caption{\footnotesize
Trajectories (red dashed curves) traced by the loaded end of the rod during a monotonic loading, for  stiffness ratios $q=0.3$ (upper part) and $q=1.2$ (lower part),
with $q=16k\lbar^{3}/(27\pi^{2}B)$. Intermediate deformed configurations (blue curves) are displayed for specific normalized loads $p=P/(k\lbar)$.
The behaviour of the imperfect ($\alpha=26.78^\circ$, right) system
shows that \lq asymptotic self-restabilization' can occur only when the perfect system ($\alpha=0$, left) does not suffer buckling.
}
\label{traiettroie}
  \end{center}
\end{figure}

The imperfect system displays two different behaviours. In the first one, self-restabilization does not occur (Fig. \ref{traiettroie}, lower part, on the right), so that the elastic rod is bent and, at increasing load, is progressively ejected from the sliding sleeve. In the second behaviour, asymptotic self-restabilization occurs (Fig. \ref{traiettroie}, upper part, on the right),
so that the rod deflection initially increases and then progressively decreases until the rod is totally inserted into the sliding sleeve. As Fig. \ref{traiettroie} shows, each of the two behaviours is connected to what is observed in the perfect system. In particular,
the asymptotic restabilization of the imperfect system occurs when the perfect system does not buckle (Fig. \ref{traiettroie}, upper part, on the left) while, oppositely, it does not occur when the perfect system does buckle at increasing load (Fig. \ref{traiettroie}, lower part, on the left).

\begin{figure*}[tp]
  \begin{center}
      \includegraphics[width= 17.5 cm]{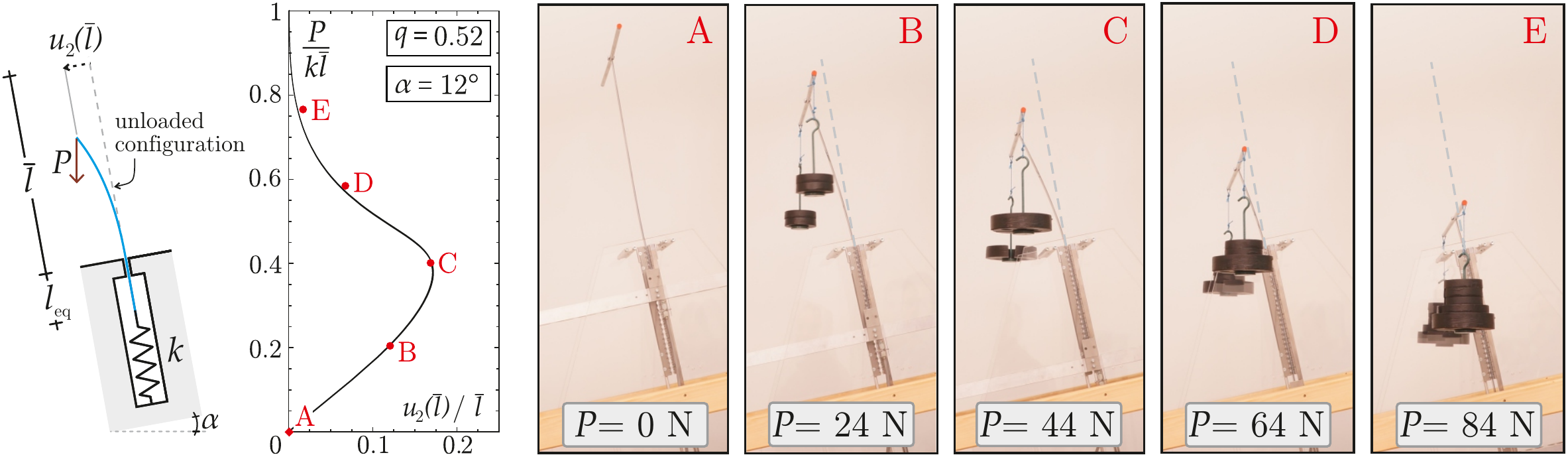}
\caption{\footnotesize
Theoretical path (line) for the dimensionless load parameter $p=P/k\lbar$
as a function of the dimensionless deflection  $u_2\left(\lbar\right)/\lbar$ for a tilt angle $\alpha=12^\circ$ and stiffness ratio $q=0.52$, with $q=16k\lbar^{3}/(27\pi^{2}B)$.
The five snapshots  A -- E reveal an initial increase followed by   a successive decrease in the deflection, when the applied vertical dead load $P$ is increased.
Each loading state A -- E is marked with a red spot in the $p - u_2\left(\lbar\right)/\lbar$ plane, so that an excellent  agreement with the theoretically predicted equilibrium path is found.
}
\label{exp1angolo}
  \end{center}
\end{figure*}

Since the effect of the imperfection is found to be crucial in the restabilization process, the mechanical behaviour of the system is investigated through both a theoretical and an experimental approach, when the inclination angle $\alpha$ is varied. In particular, the experiment reported in Fig.~\ref{exp1angolo} and performed with an inclination $\alpha=12^\circ$ shows (see the snapshots taken at $P$=$\{0, 24,44,64,84\}$N) the asymptotic self-restabilization during the application of an increasing load $P$ at the rod's free edge.
Indeed, at increasing loads, the lateral deflection of the rod initially increases (from snapshot A to C), but later decreases (from snapshot C to E) and eventually vanishes when the rod is completely inserted into the sliding sleeve, which occurs when $P\rightarrow k\lbar\cos\alpha$.

The restabilization phenomenon is related to the action of a configurational force $F_{\textup{c}}$ in the system, developed at the frictionless sliding sleeve ($s=\lin$), whenever the rod curvature does not vanish there.
Configurational forces have been introduced by Eshelby
~\cite{Eshelby:1951}~\cite{Eshelby:1956} to describe the motion of inhomogeneities within solids and have been recently interpreted as the resultant of Newtonian contact forces on a moving  inhomogeneity
~\cite{Royer-Ballarini:2016}. Essentially, a configurational force is generated whenever an elastic system can change its configuration through a release of potential energy. For example, when a dislocation is present within a stressed crystal, it tends to move within it and this tendency can be quantified with a decrease of potential energy.
For a dislocation loop represented by a closed curve of unit tangent $\tau_i(s)$ (at the point singled-out by the arc-length $s$), Burgers vector $b_j$, and stressed by $\sigma_{kj}(s)$,
the configurational force $F_h$ driving the dislocation motion is given by the Peach-Koehler relation
$$
F_h = e_{hki}\sigma_{kj}(s)b_j\tau_i,
$$
where $e_{hki}$ is the alternating Levi-Civita tensor. Recently, configurational (or \lq Eshelby-like') forces acting on elastic structures have been theoretically and experimentally proven  \cite{Bigoni-Bosi-DalCorso-Misseroni-2:2014,inietto}, exploited for the realization of self-encapsulation \cite{incapsulo} and for the design of innovative devices, such as the \lq elastica arm scale' ~\cite{Bosi-DalCorso-Misseroni-Bigoni:2014} and
the \lq torsional gun'~\cite{torsionallocomotion}.

The present analysis showing self-restabilization is based on a variational approach which provides the equations governing the dynamics of the system and gives full evidence to the configurational force. Asymptotic self-restabilization is demonstrated through the determination of the equilibrium paths of the structure under quasi-static conditions. Finally, the stability of the equilibrium paths is verified through the investigation of the response of the structure to dynamic perturbations. The theoretical predictions are fully validated by experimental tests performed on a proof-of-concept structure.

\section{The dynamics of the system from the principle of least action}\label{sec:model}
%
%
%
%
%
The equations governing the dynamics of the elastic system sketched in Fig. \ref{systemrist}
are derived below by means of the principle of the least action.
An inextensible elastic rod of length $\lbar+\hat{l}$, straight in the undeformed configuration, and with a linear elastic behaviour relating the bending moment and the curvature through a constant bending stiffness $B$, is constrained by a frictionless sliding sleeve ending with a linear spring of stiffness $k$. The sliding sleeve is tilted by an angle $\alpha$ from the vertical.
Denoting by $s$ the curvilinear coordinate along the rod, $s\in[-\hat{l},\lbar]$, the exit point of the sliding sleeve lies at $s=\lin(t)$ for all time $t$.
Therefore, the rod is partially inserted into the sleeve with the region from $s=-\hat{l}$ to $s=\lin$ lying inside the sleeve, so that the region from $s=\lin$ to $s=\lbar$ remains outside the sleeve.
The origin of the fixed $(\bm{e_x},\bm{e_y})$ frame is placed at the exit point of the sliding sleeve ($s=\lin$).
A vertical dead load $P$ is hanged at the rod's edge $s=\lbar$, so that $\bm{P}= P(-  \cos \alpha \; \bm{e_x} + \sin \alpha \; \bm{e_y})$.
In the unloaded state ($P=0$), the rod is straight and lies along the $\bm{e_x}$ axis, $\lin=0$, so that
the region from $s=0$ to $s=\lbar$ is free-standing.

Introducing the positions $x(s,t)$ and $y(s,t)$ of the rod in the reference frame $(\bm{e_x},\bm{e_y})$, and the rotation $\theta(s,t)$ between the tangent to the rod and the $\bm{e_x}$ axis,
%
%
the following boundary conditions hold:
\begin{equation}
x(\lin(t),t)=0 \: , \quad y(\lin(t),t)=0 \: , \quad \theta(\lin(t),t)=0.
\label{equa:BC-exit-point}
\end{equation}
We look now at the dynamics of the rod. Neglecting rotational inertia for the rod, we consider the kinetic energy $\mathcal{T}$ of the rod given by
\begin{equation}
\mathcal{T}(x,y)= \frac{1}{2}\int_{-\hat{l}}^{\lbar} \,  \, \gamma \, [\dot{x}^2(s,t)+\dot{y}^2(s,t)]\, \d s,
\label{equa:kinetic-energy}
\end{equation}
where $\gamma$ is the constant linear mass density of the rod and $\dot{(~)} \equiv \d (~)/ \d t$. The potential energy of the system comprises
the work of the dead load
$\bm{P}$, $\mathcal{W}_P = - \bm{P} \cdot \{ x(\lbar),y(\lbar) \}$,
the extensional strain energy of the spring
$\mathcal{U}_k=(1/2) \, k \, [x(-\hat{l})+\hat{l}]^2$,
and the bending strain energy of the rod
$\mathcal{U}_B=\int_{-\hat{l}}^{\lbar} (1/2) \, B \,  {\theta'}^2 \, \d s$, where $(~)' \equiv \d (~)/ \d s$.
As the part of the rod which lies inside the sleeve is straight
\begin{equation}
\theta'(s,t)=0 \quad \forall s \in[-\hat{l},\lin) \; \; , \, \forall t
\label{equa:null-curvature-in-sleeve}
\end{equation}
the curvature term for this region is zero and the total potential energy $\mathcal{V}$ reads
\begin{align}
\mathcal{V}(x,y,\theta,\lin) = &\frac{1}{2} \int_{\lin}^{\lbar} \, B \, {\theta'}^2 \, \d s +
\frac{1}{2} \, k \, \, [x(-\hat{l})+\hat{l}]^2
\nonumber\\
& +P \left[ x(\lbar)\, \cos \alpha  -  y(\lbar) \, \sin \alpha \right].
\end{align}
The Lagrangian to be considered is provided by the difference $\mathcal{T}-\mathcal{V}$, under the inextensibility constraints
$x'(s)=\cos \theta(s)$ and $y'(s)=\sin \theta(s)$
\begin{align}
{\cal L}(x,y,\theta,\lin) =& \mathcal{T}(x,y)- \mathcal{V}(x,y,\theta,\lin) \nonumber\\
& -
\int_{-\hat{l}}^{\lbar} N_x \, (x'-\cos \theta) \d s   \nonumber\\
& -\int_{-\hat{l}}^{\lbar} N_y \, (y'-\sin \theta) \d s,
\end{align}
where we introduce the Lagrangian multipliers $N_x(s,t)$ and $N_y(s,t)$, which are to be identified with the $x$ and $y$ components of the internal force of the rod. The dynamics of the rod is consequently given by first-order conditions to minimize the action ${\cal A}$
\begin{align}
{\cal A}  (x,y,\theta,\lin)&= \int_{t_1}^{t_2}  {\cal L}(x(s,t),y(s,t),\theta(s,t),\lin(t))  \, \d t.
\end{align}
%
%
%
%
%
Following this minimization procedure (reported in the appendix), the equations governing the dynamics of the region
of the rod outside the sliding sleeve, $s\in(\lin(t),\lbar]$, are obtained as
\begin{subequations}
\label{sys:dynamics}
\begin{align}
x'(s,t) = \cos \theta \; &, \quad y'(s,t)  = \sin \theta \, , \\
B \, \theta'(s,t)  = M \; &, \quad M'(s,t)  = N_x \, \sin \theta - N_y \, \cos \theta \, , \\
N_x'(s,t)  = \gamma \, \ddot{x} \; &, \quad N_y'(s,t)  = \gamma \, \ddot{y} \, ,
\end{align}
\end{subequations}
complemented by the following boundary conditions
\begin{subequations}
\label{sys:dynamics-bcnd}
\begin{align}
& x(\lin(t),t)    = 0  \, , \; y(\lin(t),t) = 0 \, , \; \theta(\lin(t),t) = 0 \, , \\
& N_x(\lin(t),t)   = -F_{\text{c}} - k \lin(t) - \gamma (\lin+\hat{l})\ddot{\lin}(t)  \, , \\
& M(\lbar,t)     = 0  \, , \; N_x(\lbar,t) = -P \cos \alpha \, , \; N_y(\lbar,t) = P \sin \alpha \, .
\end{align}
\end{subequations}
The action of the configurational force $F_{\text{c}}$  generated at the sliding sleeve exit
\cite{Bigoni-Bosi-DalCorso-Misseroni-2:2014}, \cite{inietto}, \cite{incapsulo}, \cite{Bosi-DalCorso-Misseroni-Bigoni:2014},
\begin{equation}
F_{\text{c}}=\frac{M^2(\lin(t),t)}{2B},
\label{config}
\end{equation}
is disclosed from the equilibrium equation along the $x$-axis (\ref{sys:dynamics-bcnd}b),
recalling that $M(\lin)=M(\lin^+)$.


%
%
%
%
%
%
\section{Asymptotic self-restabilization}\label{sectangolo}
%
%
%
%
%

Considering the quasi-static condition $\ddot{x} = \ddot{y} = \ddot{\lin} = 0$ in
equations (\ref{sys:dynamics})-(\ref{sys:dynamics-bcnd}),
the equilibrium configuration  $\bm{w}_{\text{eq}}=(x_{\text{eq}}(s),y_{\text{eq}}(s),\theta_{\text{eq}}(s),l_{\text{eq}})$
for a fixed $P$  is governed by the following equations for
$s\in ( l_{\textup{eq}},\lbar \, ]$
\begin{equation}
\label{sys:equilibrium}
\begin{array}{lll}
B \, \theta_{\text{eq}}''(s)  = {N_x}_{\text{eq}}(s) \, \sin \theta_{\text{eq}}(s) - {N_y}_{\text{eq}}(s) \, \cos \theta_{\text{eq}}(s), \\
{N_x}_{\text{eq}}'(s)  =0,  \\
{N_y}_{\text{eq}}'(s)  = 0,
\end{array}
\end{equation}
subject to the following boundary conditions
\begin{align}
\label{sys:equilibrium-bcnd}
& x_{\text{eq}}(\leqq)     = 0,  \quad
y_{\text{eq}}(\leqq) = 0, \quad
\theta_{\text{eq}}(\leqq) = 0,\nonumber \\
& \frac{M_{\text{eq}}^2(\leqq)}{2B}   =  P \cos \alpha- k \, \leqq, \nonumber  \\
&{N_x}_{\text{eq}}(\leqq) = -P \cos \alpha, \quad {N_y}_{\text{eq}}(\leqq) = P \sin \alpha.
\end{align}

The governing equations (\ref{sys:equilibrium}) and the boundary conditions (\ref{sys:equilibrium-bcnd})
can be reduced to the rotational equilibrium given by a differential equation for the rotation field at equilibrium $\theta_{\text{eq}}$
\begin{align}
\label{diffproblemangolo}
\theta_{\textup{eq}}^{''}(s)+\lambda^2\left[\cos\alpha\sin\theta_{\textup{eq}}(s)+\sin\alpha\cos\theta_{\textup{eq}}(s)\right]=0,
\nonumber\\
 s\in (l_{\textup{eq}},\lbar \, ],
\end{align}
with the moving boundary $\leqq$,  defined by the axial equilibrium at the exit of the sliding sleeve,
\begin{align}
\label{diffproblemangolo2}
l_{\textup{eq}}=\dfrac{B}{k}\left(\lambda^2\cos\alpha-\dfrac{1}{2}\left[\theta_{\textup{eq}}^{'}(l_{\textup{eq}})\right]^{2}\right),
\end{align}
where the load parameter $\lambda^2=P/B$ has been introduced and the rotation field $\theta_{\textup{eq}}(s)$ is subject to the boundary conditions
$\theta_{\textup{eq}}(l_{\textup{eq}})=0$ and
$\theta_{\textup{eq}}^{'}\left(\lbar\right)=0$.

\subsection{Equilibrium paths}
%
%
%
%
%
The solution at equilibrium can be obtained through an analytical manipulation
of equations ~\eqref{diffproblemangolo}, ~\eqref{diffproblemangolo2}, based on change of variables and integrations.
Defining   $\theta_{\lbar}$ as the rotation at the free end ($s=\lbar$) at equilibrium,
$\theta_{\textup{eq}}(\lbar)=\theta_{\lbar}$,
the dimensionless load $p$ and the dimensionless stiffness ratio $q$ as
\begin{equation}
p=\dfrac{P}{k\lbar} \: , \qquad q=\dfrac{16k\lbar^{3}}{27\pi^{2}B},
\end{equation}
the equilibrium configuration (restricted to the first deformation mode)
is provided at varying load $p$ and tilt angle $\alpha$  as the solution $\theta_{\lbar}(p)$ and $l_{\textup{eq}}(p)$ to the following system of nonlinear equations
\begin{equation}\begin{array}{lll}
\label{cubica}
(1-2\eta^2)^2p^{3}-2(1-2\eta^2)p^{2}+p=
\dfrac{16\left[\mathcal{K}\left(\eta\right)-\mathcal{K}\left(m,\eta\right)\right]^2}{27 \pi^2
q},\\[3 mm]
l_{\textup{eq}}=p(1-2\eta^2)\lbar.
\end{array}
\end{equation}
In equation (\ref{cubica}),   $\mathcal{K}\left(\eta\right)$ and $\mathcal{K}\left(m,\eta\right)$ are the complete and incomplete elliptic integral
of the first kind respectively and
the parameters $m$ and $\eta$ depend on the angles $\alpha$ and $\theta_{\lbar}$ as
\begin{equation}
m=\arcsin\left[\frac{\sin\frac{\alpha}{2}}{\eta}\right],
\qquad
\eta=\sin\frac{\theta_{\lbar}+\alpha}{2}.
\end{equation}

Once the nonlinear system (\ref{cubica}) is solved, the kinematical fields can be evaluated through

\begin{equation}
\begin{array}{lll}
 \theta_{\textup{eq}}(s)=&2\arcsin\left[\eta\,\textup{sn}\left(\lambda\left(s-l_{\textup{eq}}\right)+\mathcal{K}\left(m,\eta\right),\eta\right)\right]-\alpha,\\[4 mm]
u_{1}(s) =&-\frac{2\eta}{\lambda}\sin\alpha\Bigl\{ \textup{cn}\bigl(\lambda\left(s-l_{\textup{eq}}\right)+\mathcal{K}(m,\eta),\eta\bigr)\\
&-\textup{cn}\bigl(\mathcal{K}(m,\eta),\eta\bigr)\Bigr\} -l_{\textup{eq}}+ \cos\alpha\Bigl\{ -s\\
&+\frac{2}{\lambda}\Bigl[E\bigl[\textup{am}\bigl(\lambda\left(s-l_{\textup{eq}}\right)+\mathcal{K}(m,\eta),\eta\bigr)\bigr]\\
&-E\bigl[\textup{am}\bigl(\mathcal{K}(m,\eta),\eta\bigr)\bigr]\Bigr]\Bigr\},\\[4 mm]
 u_{2}(s) =& \frac{2\eta}{\lambda}\cos\alpha\Bigl\{ \textup{cn}\bigl(\lambda\left(s-l_{\textup{eq}}\right)+\mathcal{K}(m,\eta),\eta\bigr)\\
&-\textup{cn}\bigl(\mathcal{K}(m,\eta),\eta\bigr)\Bigr\} +\sin\alpha\Bigl\{ -s\\
&+\frac{2}{\lambda}\Bigl[E\bigl[\textup{am}\bigl(\lambda\left(s-l_{\textup{eq}}\right)+\mathcal{K}(m,\eta),\eta\bigr)\bigr]\\
&-E\bigl[\textup{am}\bigl(\mathcal{K}(m,\eta),\eta\bigr)\bigr]\Bigr]\Bigr\} ,
\end{array}
\label{elasticablade}
\end{equation}

where $u_{1}(s)$, $u_{2}(s)$ are the axial and transverse displacement fields,
the functions am, cn and sn denote the Jacobi amplitude, Jacobi cosine amplitude and Jacobi sine amplitude functions,
 while $E(x,\eta)$ is the incomplete elliptic integral of the second kind of modulus $\eta$.
 Equations~\eqref{elasticablade}$_2$ and ~\eqref{elasticablade}$_3$, evaluated at the rod's loaded end, provide the following expressions for the axial and transversal
 displacements of the free edge of the rod
\begin{align}\label{spostamentifinaliangolo}
u_{1}\left(\lbar\right)&=\frac{1}{\lambda}\Bigl\{ 2\eta\sin\alpha\cos m+\cos\alpha\bigl[2E\left(\eta\right)-2E\left(m,\eta\right) \nonumber\\
&+\mathcal{K}\left(m,\eta\right)-\mathcal{K}\left(\eta\right)\bigr]+\mathcal{K}\left(m,\eta\right)-\mathcal{K}\left(\eta\right)\Bigr\} -l_{eq},\nonumber\\[4 mm]
u_{2}\left(\lbar\right)&=\frac{1}{\lambda}\Bigl\{ 2\eta\cos\alpha\cos m+\sin\alpha\bigl[2E\left(m,\eta\right)-2E\left(\eta\right)
\nonumber\\
&+
\mathcal{K}\left(\eta\right)-\mathcal{K}\left(m,\eta\right)\bigr]\Bigr\}.
\end{align}

The solution of the cubic equation (\ref{cubica})$_1$ provides
three  values of $p$ associated to the same triad given by the free end rotation $\theta_{\lbar}$,
the stiffness ratio $q$ and the angle $\alpha$, namely
\begin{equation}
p^{\mathcal{A}}=p^{\mathcal{A}}(\theta_{\lbar},q,\alpha),\quad
p^{\mathcal{B}}=p^{\mathcal{B}}(\theta_{\lbar},q,\alpha),\quad
p^{\mathcal{C}}=p^{\mathcal{C}}(\theta_{\lbar},q,\alpha),
\end{equation}
with $p^{\mathcal{C}}$ being always real, while $p^{\mathcal{A}}$ and $p^{\mathcal{B}}$ taking real or complex values depending on the positive or negative sign of the discriminant $\Delta$,
\begin{align}
\label{discriminant}
    \Delta(\theta_{\lbar}, q, \alpha)=\frac{64}{27(1-2\eta^2)^5}\frac{\left[\mathcal{K}\left(\eta\right)-\mathcal{K}\left(m,\eta\right)\right]^2}{\pi^2 q}\times\nonumber\\
    \left[1-(1-2\eta^2)\frac{4\left[\mathcal{K}\left(\eta\right)-\mathcal{K}\left(m,\eta\right)\right]^2}{\pi^2 q}\right].
\end{align}
When the discriminant $\Delta$ is positive, the roots $p^{\mathcal{A}}, p^{\mathcal{B}}, p^{\mathcal{C}}$ are all real and can be expressed through the following trigonometric relation \cite{Cox2012Galois-Theory}

\begin{widetext}
\begin{equation}
\label{viete}
\left\{
\begin{array}{cc}
p^{\mathcal{A}}\\
p^{\mathcal{B}}\\
p^{\mathcal{C}}
\end{array}
\right\}
=\frac{2
\left\{
1+\mbox{sign}[1-2\eta^2]\cos\left[
    \frac{1}{3}\left(2\pi(j-1) +
    \arccos\left[\left(
    \frac{8(1-2\eta^2)\left[\mathcal{K}\left(\eta\right)-\mathcal{K}\left(m,\eta\right)\right]^2}{\pi^2 q}-1\right)\mbox{sign}\left[1-2\eta^2\right]
    \right)
    \right]
    \right]
\right\}
}{3(1-2\eta^2)}
\end{equation}
\end{widetext}
where $j=\left\{2,3,1\right\}$, respectively for the root $\left\{p^\mathcal{A},p^\mathcal{B},p^\mathcal{C}\right\}$.

\begin{figure*}[tp]
  \begin{center}
      \includegraphics[width= 17 cm]{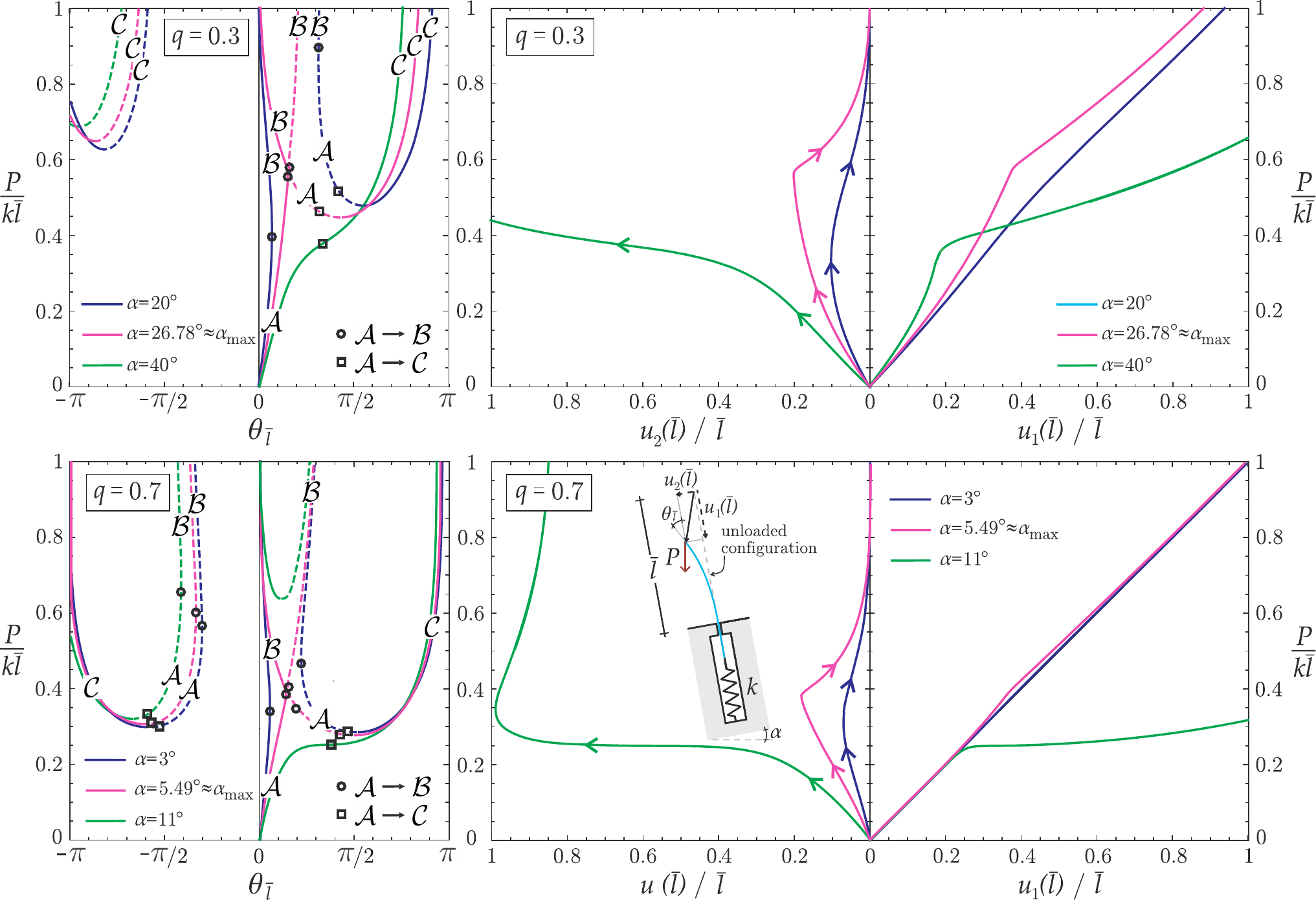}
\caption{\footnotesize Equilibrium paths for the system with stiffness ratio $q=0.3$ (upper part) and $q=0.7$ (lower part) and for three different tilt angles.
In particular, the equilibrium paths are shown for angles $\alpha$ smaller than  (blue curves), corresponding to (pink curves), and higher than (green curves) the maximum tilt angle $\alpha_{\textup{max}}$ for which  asymptotic self-restabilization is observed. Therefore, in two cases ($\alpha=40^\circ$ and $\alpha=11^\circ$, green curves)
restabilization does not occur.
Left: Roots $p^\mathcal{A}$, $p^\mathcal{B}$, and  $p^\mathcal{C}$   as a function of the free end angle $\theta_{\lbar}$ at equilibrium.  Stable and unstable paths (as found in Sect. \ref{stabmeth}) are depicted through continuous and dashed lines, respectively. The transition point in the  equilibrium paths is marked through a circle (for the connection of  $\mathcal{A}$ with $\mathcal{B}$) and a square (for the connection of $\mathcal{A}$ with $\mathcal{C}$). Right:
Dimensionless load $P/(k\,\lbar)$ applied to the rod's end reported as a function of its axial (right part) and transverse (left part)
displacements, $u_1(\lbar)$ and $u_2(\lbar)$, made dimensionless through division by the initial external length $\lbar$. Only the stable paths departing from the unloaded configuration and attained during a monotonic increase of the loading are reported.
The evolution of the deformed configuration for a self-restabilizing  system ($q=0.3$ and $\alpha=26.78^\circ$)
is sketched in Fig. \ref{traiettroie} (upper part, right). The configuration evolution for a non
self-restabilizing  system is qualitatively similar to that sketched  in
Fig. \ref{traiettroie} (lower part, right) for the case $q=1.2$ and $\alpha=26.78^\circ$.
}
\label{caricoangolo}
  \end{center}
\end{figure*}

Equilibrium paths for the system with stiffness ratio $q=0.3$ and $q=0.7$ are reported in Fig.~\ref{caricoangolo}  (upper and lower parts, respectively)
for three different tilt angles $\alpha$.
The roots $p^{\mathcal{A}}$, $p^{\mathcal{B}}$, and $p^{\mathcal{C}}$, given by equation (\ref{viete}), of the cubic equation (\ref{cubica})$_1$
are plotted in Fig.~\ref{caricoangolo} (left) as functions of the free end rotation $\theta_{\lbar}$. Stable and unstable paths (see next section
for the details)  are reported as continuous and dashed lines, respectively. The limit values of the ranges where the
roots $p^\mathcal{A}$, $p^\mathcal{B}$, and $p^\mathcal{C}$ prevail are marked in the figure with circular and square spots,
the former indicating the transition from solution ${\mathcal{A}}$ to ${\mathcal{B}}$, the latter from ${\mathcal{A}}$ to ${\mathcal{C}}$.

Using equation (\ref{spostamentifinaliangolo}),
the stable equilibrium paths (departing from the undeformed state) are reported in Fig.~\ref{caricoangolo} (right) in terms of
the dimensionless load $p=P/k\lbar$ as a function of the dimensionless displacement components, namely, the axial displacement
$u_1(\lbar)/\lbar$  and the deflection $u_2(\lbar)/\lbar$.

It is clear from Fig.~\ref{caricoangolo} that the equilibrium paths are strongly affected by the tilt angle $\alpha$ and, depending on
this parameter, the \textit{asymptotic self-restabilization} may or may not occur.
In particular, during a monotonic loading, whenever the tilt angle is smaller or equal than the maximum tilt angle for restabilization,
 $\alpha\leq\alpha_{max}(q)$,
the rod initially leaves the undeformed straight configuration and then \lq spontaneously' and smoothly returns back to this configuration,
until the limit is approached of complete penetration of the rod into  the sliding sleeve.
Therefore the rod deflection $u_2(\lbar)$ initially increases, but later decreases until it vanishes,
in the limit $p\rightarrow 1/\cos\alpha$, so that \textit{asymptotic self-restabilization} occurs.
When this does not occur,  $\alpha>\alpha_{max}(q)$,
the complete penetration of the rod into the sliding sleeve does never realize.
Therefore, in the cases considered in Fig.~\ref{caricoangolo}, asymptotic self-restabilization occurs for $\alpha \leq \alpha_{\textup{max}}(q=0.3)\approx26.78^\circ$  (Fig.~\ref{caricoangolo}, upper part) and   for $\alpha \leq\alpha_{\textup{max}}(q=0.7)\approx 5.49^\circ$  (Fig.~\ref{caricoangolo}, lower part).

\begin{figure}[tp]
  \begin{center}
      \includegraphics[width= 8.5 cm]{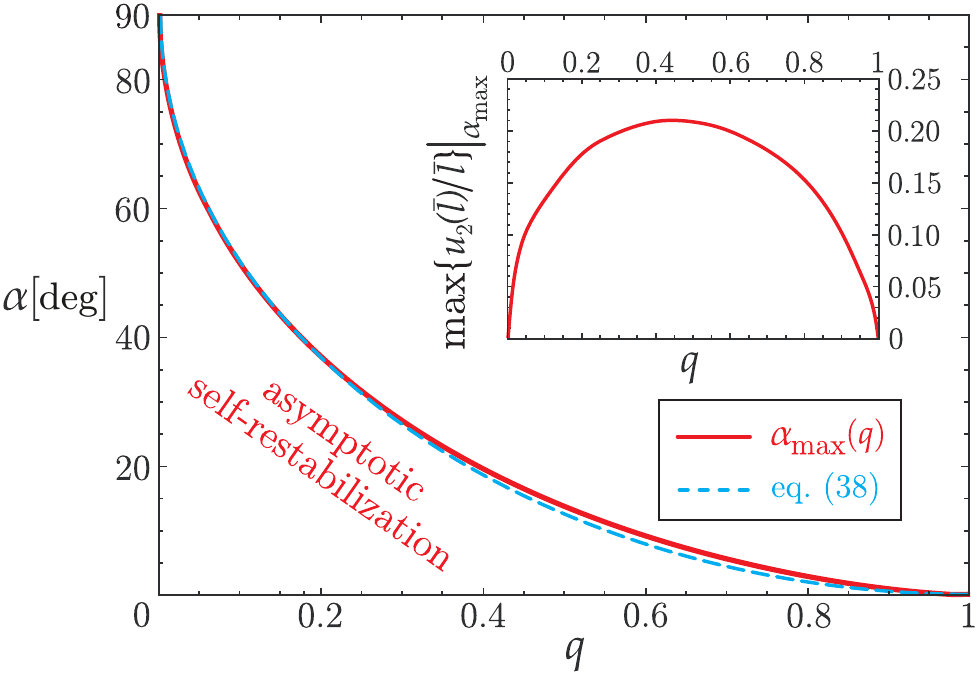}
\caption{\footnotesize
The region enclosed by red continuous line and the two axes  contains all the pairs of the stiffness
ratio $q$ and the tilt angle $\alpha$ for which asymptotic self-restabilization occurs.
The blue dashed line is the approximation (\ref{cerchio}) for
the maximum tilt angle $\alpha_{\textup{max}}$ for which  asymptotic self-restabilization
is observed, reported as red continuous line. Inset: Maximum deflection (made dimensionless through division by the initial external length $\lbar$) occurring during asymptotic self-restabilization at varying the stiffness ratio $q$, evaluated considering $\alpha_{\textup{max}}(q)$. It is observed that the maximum value of the deflection is attained for $q=0.45$, corresponding to $\alpha_{\textup{max}}=16.29^\circ$, for which  max$\left\{u_2(\lbar)/\lbar\right\}|_{\alpha_{\textup{max}}}=0.21$.
}
\label{zonaristangolo}
  \end{center}
\end{figure}

The restabilization phenomenon is displayed whenever the discriminant $\Delta$, ~\eqref{discriminant}, takes non-positive values for a set of the end rotations $\theta_{\lbar}$.
Through this criterion,  the  values of the stiffness ratio $q$ and the tilt angle $\alpha$ for which the \textit{asymptotic self-restabilization} occurs  have been numerically obtained and reported in Fig.~\ref{zonaristangolo}.
It is observed that  the maximum tilt angle $\alpha_{\textup{max}}(q)$ (reported as red continuous line in Fig.~\ref{zonaristangolo}) can be approximately described through the following equation
\begin{equation}\label{cerchio}
    \alpha_{\textup{max}}(q)\simeq\frac{\pi}{2}\left[1-\sqrt[1.97]{1-\left(1-q\right)^{1.97}}\right],
\end{equation}
which is represented as the blue dashed line in Fig.~\ref{zonaristangolo}.

We finally note that asymptotic self-restabilization may occur only for $q<1$, which is the condition for which the bifurcation does not occur in the case when  the tilt angle is null ($\alpha=0$) \cite{Bigoni-Bosi-DalCorso-Misseroni-1:2014}. Therefore, asymptotic self-restabilization is strictly related to the fact that the perfect system does not display buckling, but
only a pure penetration of the straight rod into the sliding sleeve at increasing load.
As a further evidence of this concept, trajectories of the loaded end, together with intermediate deformed configurations, obtained during a monotonic loading are shown in Fig. \ref{traiettroie} for $q$=0.3 (upper part) and for $q$=1.2 (lower part) in the perfect ($\alpha=0$, left) and imperfect case ($\alpha=26.78^\circ$, right).

Finally, as a quantitative measure of the self-restabilization effect, the maximum deflection $u_2(\lbar)$ attained during loading has been evaluated for a tilt angle corresponding to the maximum tilt angle,  $\alpha=\alpha_{\textup{max}}$. This quantity  is reported as a function of the stiffness ratio $q$ in the inset of Fig.~\ref{zonaristangolo}. It is observed that the largest  deflection which may occur in the system during the asymptotic self-restabilization is about $21\%$ of the initial external length $\lbar$ and is attained for $q\approx 0.45$, which corresponds to $\alpha_{\textup{max}}\approx16.29^\circ$ as maximum tilt angle.

\subsection{Stability from vibrations}\label{stabmeth}
\begin{figure}[tp]
  \begin{center}
      \includegraphics[width= 8.5 cm]{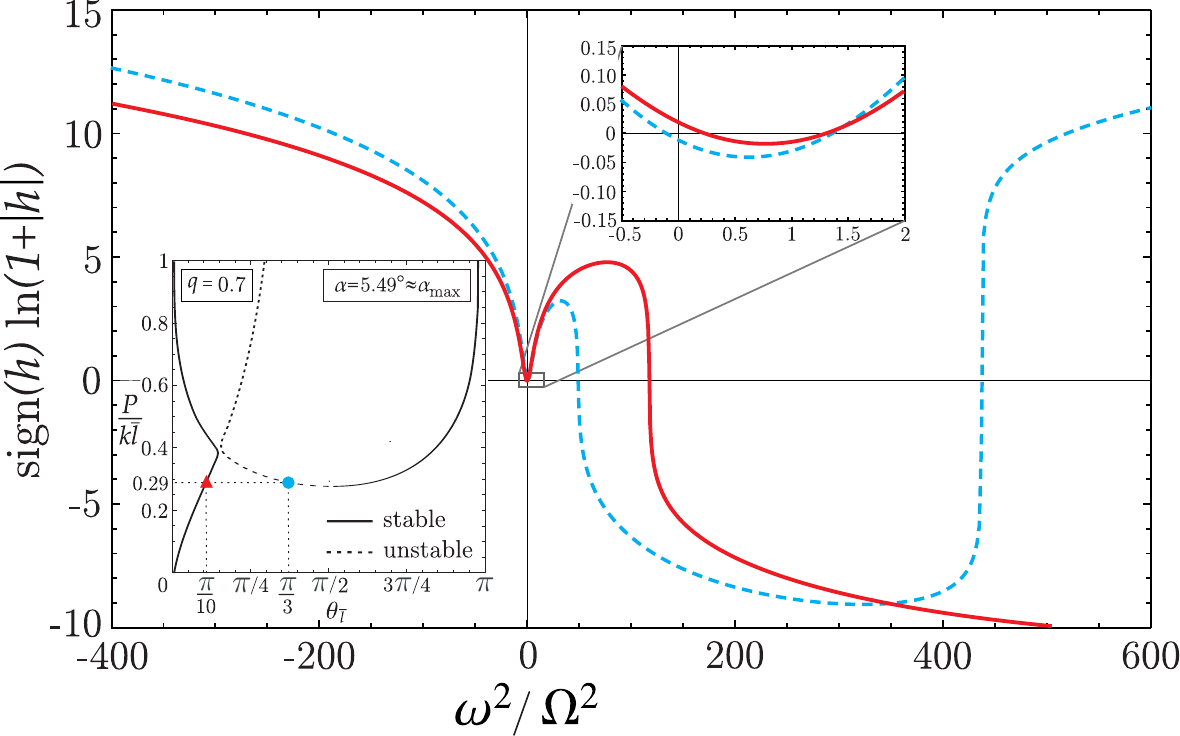}
\caption{ The stability of each equilibrium configuration of the mechanical system showing restabilization is detected by finding the roots $\omega^2_i$ of the determinant $h(\omega^2_i)$ and revealed by plotting the
quantity $\text{sign}(h) \, \ln(1+|h|)$.
Here two equilibrium configurations are considered,  both characterized by the same values $q=0.7$, $\alpha=5.49^\circ$,  and $P/k \lbar \simeq 0.29$, but  different  free end rotations
$\theta_{\lbar}=0.1\pi$ and $\theta_{\lbar}=\pi/3$,  represented respectively as a red triangle and a blue dot in the inset.
In the former case (solid red curve), the lowest three roots are given by
$\omega_1^2 \simeq 0.24 \, \Omega^2$,
$\omega_2^2 \simeq 1.36 \, \Omega^2$,  and
$\omega_3^2 \simeq 123.56 \, \Omega^2$, with $\Omega=(2 \pi / {\lbar}^2 ) \, \sqrt{B/\gamma}$, while in the latter case (dashed blue curve), they are
$\omega_1^2 \simeq -0.11 \, \Omega^2$,
$\omega_2^2 \simeq 1.36 \, \Omega^2$, and
$\omega_3^2 \simeq 49.1 \, \Omega^2$. As a result, the configuration with $\theta_{\lbar}=0.1\pi$ is stable (with a positive $\omega_1^2$)
while the configuration with $\theta_{\lbar}=\pi/3$ is unstable (with a negative $\omega_1^2$).
}
\label{fig:detH}
  \end{center}
\end{figure}

%
%
%
Once the equilibrium configurations are found by solving  equations (\ref{sys:equilibrium})-(\ref{sys:equilibrium-bcnd}), we test the stability by computing
the linear vibrations of the system about each equilibrium configuration
\begin{align}
\textbf{W}_{\text{eq}}(s,t)=\Big(x_{\text{eq}}(s,t), y_{\text{eq}}(s,t), \theta_{\text{eq}}(s,t), l_{\text{eq}}(t),\nonumber\\
M_{\text{eq}}(s,t), {N_x}_\text{eq}(s,t), {N_y}_\text{eq}(s,t)\Big),
\end{align}
so that we consider the dynamics
equations (\ref{sys:dynamics})-(\ref{sys:dynamics-bcnd}) and look for the linear modes $\checkVIB{\textbf{W}}(s)$
defining the perturbed configuration as follows
\begin{equation}
\label{sys:vib-modes}
\textbf{W}(s,t) = \textbf{W}_{\text{eq}}(s,t) + \epsilon \, \checkVIB{\textbf{W}}(s) \, \cos \omega t,
\end{equation}
where $\omega$ is the frequency and  $|\epsilon| \ll 1$.
Inserting the representation (\ref{sys:vib-modes}) into the dynamics equations, using equations (\ref{sys:equilibrium})
and retaining only the linear terms in $\epsilon$, yields the following non-autonomous linear system of differential equations for the modes
\begin{equation}
\begin{array}{llll}
\label{sys:vibrations}
\checkVIB{x}'(s) =& - \checkVIB{\theta} \, \sin \theta_{\text{eq}} \, , \;
\checkVIB{y}'(s) =  \checkVIB{\theta} \, \cos \theta_{\text{eq}} \, , \;
B \, \checkVIB{\theta}'  = \checkVIB{M}, \\[3mm]
\checkVIB{{N}}_x'(s)  =& - \omega^2 \gamma\, \checkVIB{x} \: , \;
 \checkVIB{{N}}_y'(s)  = - \omega^2 \gamma\, \checkVIB{y} \, ,
\\ [3mm]
\checkVIB{M}'(s)  =& \checkVIB{N}_x \, \sin \theta_{\text{eq}} - \checkVIB{N}_y \, \cos \theta_{\text{eq}} \\[3mm]
&+ \checkVIB{\theta} \, ({N_x}_{\text{eq}} \, \cos \theta_{\text{eq}} + {N_y}_{\text{eq}} \, \sin \theta_{\text{eq}}  ).
\end{array}
\end{equation}
The boundary conditions for the modes are found by inserting equation (\ref{sys:vib-modes}) in
equations (\ref{equa:BC-exit-point}),
(\ref{sys:dynamics-bcnd}),
and using equation (\ref{sys:equilibrium-bcnd}), to obtain
\begin{subequations}
\label{sys:modes-bcnd}
\begin{equation}
\checkVIB{x}(\leqq)     = -\checkVIB{\lin} \, ,  \; \checkVIB{y}(\leqq) = 0 \, , \;
\checkVIB{\theta}(\leqq) = -\checkVIB{\lin} \, \frac{M_{\text{eq}}(\leqq)}{B}, \label{equa:modes-BC-start}
\end{equation}
\begin{align}\label{equa:modes-BC-start-bis}
\checkVIB{N}_x(\leqq)  =&
-\frac{M(\leqq) \, \checkVIB{M}(\leqq)}{B}
+\lambda^2 \sin \alpha\,M(\leqq) \, \checkVIB{\lin}
- k \checkVIB{\lin}\nonumber\\
&+ \gamma \,  \omega^2 \, (\leqq+\hat{l}) \, \checkVIB{\lin},
\end{align}
\begin{equation}
\checkVIB{M}(\lbar)    = 0 \, , \; \checkVIB{N}_x(\lbar) = 0 \, , \; \checkVIB{N}_y(\lbar) = 0.
\label{equa:modes-BC-end}
\end{equation}
\end{subequations}
The stability of the equilibrium solution $\textbf{W}_{\text{eq}}(s,t)$ is therefore
related to the sign of $\omega^2$, with $\omega^2 > 0$ corresponding to a stable configuration,
and $\omega^2 < 0$ to an unstable configuration.
We solve the linear boundary value problem (\ref{sys:vibrations})-(\ref{sys:modes-bcnd}) using a shooting method approach:
at $s=\leqq$ only three quantities are unknown $\bm{X}=(\checkVIB{\lin},\checkVIB{M}(\leqq),\checkVIB{N}_y(\leqq))$,
the other being given by equations (\ref{equa:modes-BC-start})-(\ref{equa:modes-BC-start-bis}).
Once the integration of the linear system (\ref{sys:vibrations}) is performed, the
boundary conditions (\ref{equa:modes-BC-end}) depend linearly on $\bm{X}$, so that writing
$\bm{Y}=(\checkVIB{M}(\lbar), \checkVIB{N}_x(\lbar),\checkVIB{N}_y(\lbar))$
this linear problem can be expressed as
\begin{equation}
\bm{Y} = \bm{H}(\omega^2) \, \bm{X} =  \bm{0}.
\end{equation}
Requiring $\bm{X}$ to be non-zero imposes the matrix $\bm{H}$ to be singular. Defining $h(\omega^2) = \det[\bm{H(}\omega^2)]$, the mode frequencies $\omega_i$ can be obtained as the roots of  $h$, namely, solving the equation $h(\omega^2_i)=0$.
As equations (\ref{sys:vibrations})-(\ref{sys:modes-bcnd}) can be recast as a Sturm-Liouville problem \cite{Bigoni-Bosi-DalCorso-Misseroni-1:2014}, we know that the lowest eigenvalue $\omega^2_1$ is finite \cite{Boyce1959Vibrations-Of-Twisted}, \cite{Rao2007Vibration-of-continuous}.
For each equilibrium solution, we plot $\mbox{sign}(h)\ln(1+|h|)$ as function of $\omega^2$ and record the smallest
eigenvalue $\omega_1^2$. The equilibrium is then stable if the lowest eigenvalue is strictly positive, and unstable if it is strictly negative.
Such analysis allowed the definition of stability for the equilibrium paths reported in Figs. \ref{caricoangolo}. As an example, the typical behaviour of the
 function $\mbox{sign}(h)\ln(1+|h|)$ for a stable and an  unstable equilibrium configuration is reported in Figure~\ref{fig:detH}, showing respectively a positive and a negative value for the lowest  root $\omega_1^2$ of the determinant $h(\omega^2)$. The two considered equilibrium configurations are
 both characterized by the same values of $q=0.7$, $\alpha=5.49^\circ$,  and $P/k \lbar \simeq 0.29$, while they correspond to the two different free end rotations,
namely, $\theta_{\lbar}=0.099\pi$ (stable configuration) and $\theta_{\lbar}=\pi/3$ (unstable configuration).

%
%
%
%
%
\subsection{Experiments}
\begin{figure}[tp]
  \begin{center}
      \includegraphics[width= 8.5 cm]{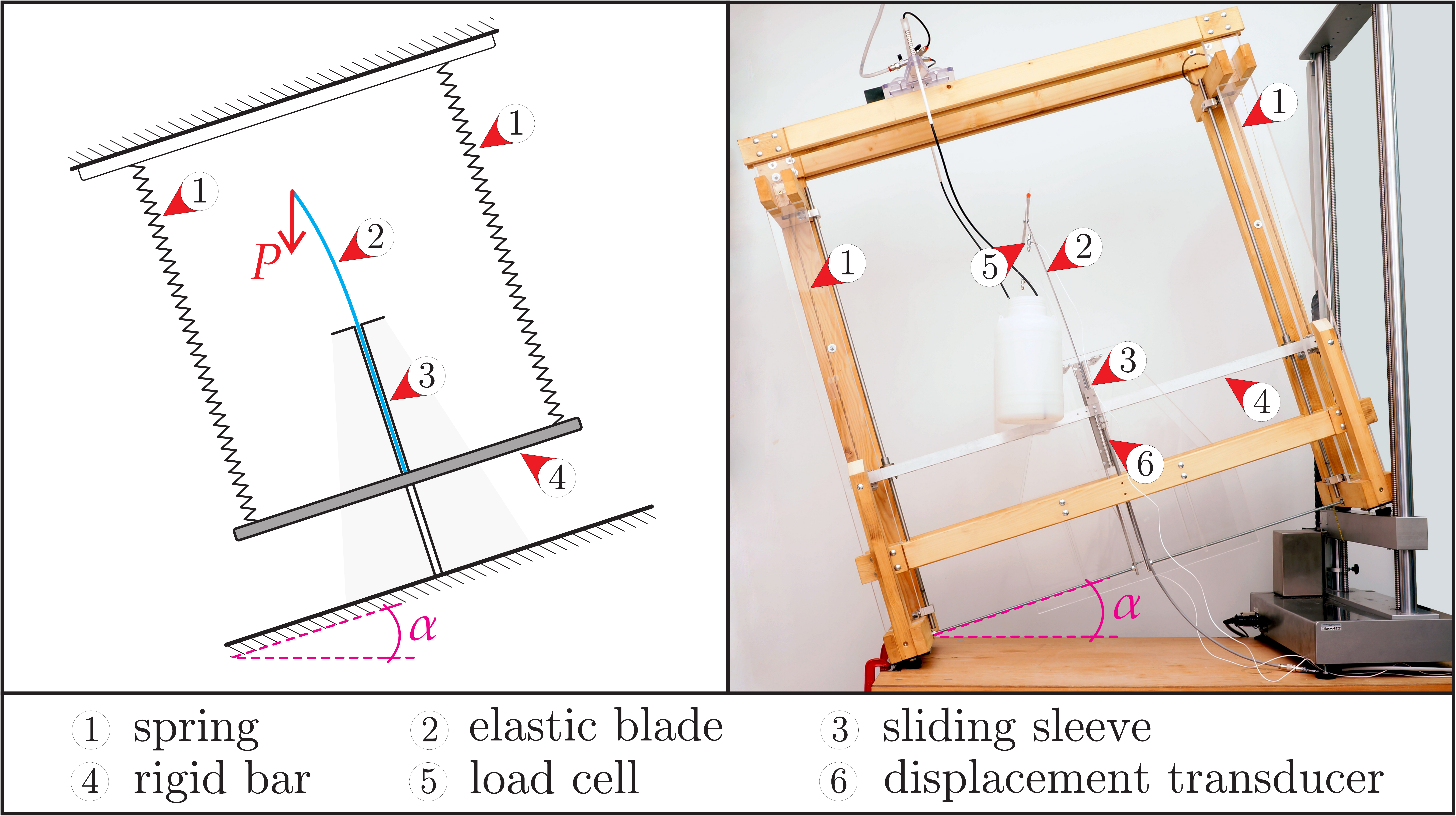}
\caption{\footnotesize The design scheme (left) employed to realize the structure shown in Fig.~\ref{systemrist} and its practical realization (right) used in the experimental validation of the asymptotic self-restabilization.}
\label{prototipo}
  \end{center}
\end{figure}

%
%
%
%
Experiments have been performed at the \lq Instabilities Lab' of the University of Trento on the prototype reported in Fig.~\ref{prototipo} as a sketch and a photo, where the linear elastic axial spring  in Fig.~\ref{systemrist} has been realized by hanging a highly-stiff bar (to which the elastic rod is clamped and orthogonal) to two carbon steel (EN 10270-1 SH) springs (D19100, 1.25 mm wire diameter and 8 mm mean coil diameter, $k=225$ N/m, purchased from D.I.M.). The stiff bar can only rigidly translate as constrained by two linear bushings (LHFRD12, Misumi Europe) parallel to the rod in its undeformed state. The tilt angle  $\alpha$ has been provided by simply inclining the prototype, through a lifting the right support (with the movable crosshead of a MIDI 10 load frame, from Messphysik).  The penetration length $l_{\textup{eq}}$ of the rod has been obtained by measuring the displacement of the lower edge of the rod through a magnetic non-contact displacement transducer GC-MK5 (from Gemac). The data have been acquired with a NI CompactDAQ system, interfaced with Labview 8.5.1 (National Instruments).
All the photos were
taken with a Sony NEX 5N digital camera, equipped with 3.5-5.6/18-55 lens (optical steady shot from Sony Corporation).

 A first experiment showing the asymptotic self-restabilization is reported in Fig. \ref{exp1angolo}, where five snapshots of the system for a tilt angle $\alpha=12^\circ$ are reported at increasing  applied loading, $P=\left\{0,24,44,64,84\right\}$ N, on a C62 Carbon-steel rod (25 mm x 2 mm cross section, bending stiffness $B=2.70$ Nm$^2$) of external length in the unloaded configuration $\bar{l}=0.47$ m. It can be noted that after an initial increase in the deflection (from snapshot A to C), a further increase of load realizes a decrease in the deflection (from snapshot C to E). The experimental values of the deflection  (normalized through division of the initial external length), measured from image post-processing, are also reported dotted, as function of the dimensionless load and compared with the theoretical equilibrium path.

\begin{figure}[tp]
  \begin{center}
      \includegraphics[width= 8.5 cm]{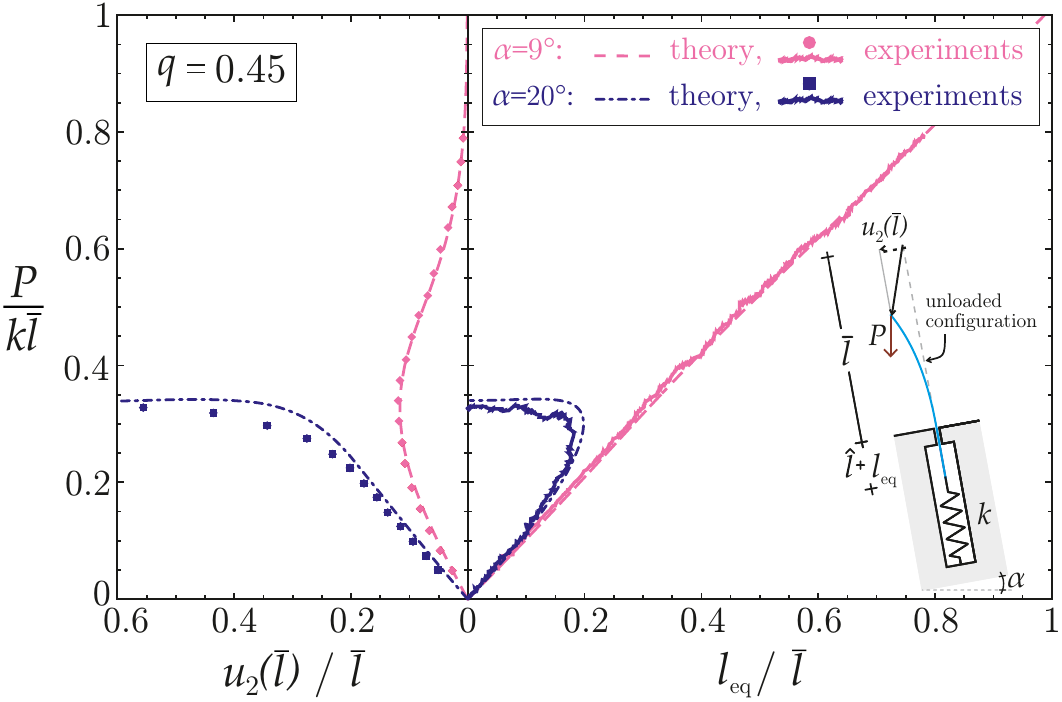}
\caption{\footnotesize Comparison between theoretical and experimental results. The dimensionless load $p=P/(k\lbar)$ is reported as function of both the dimensionless deflection $u_2(\lbar)/\lbar$ (left) and amount of the rod penetrated into the sliding sleeve $l_{\textup{eq}}/\lbar$ (right). The behaviour of the deflection reveals that for $\alpha=9^\circ$
(shown pink) the asymptotic self-restabilization does occur, while for $\alpha=20^\circ$ (shown blue) the asymptotic self-restabilization does not. Continuous and dashed lines denote respectively experimental measures and theoretical predictions.
Dots and squares indicate experimental measures extracted from a video of the experiment through image post-processing,
for $\alpha=9^\circ$ and $\alpha=12^\circ$ respectively.}
\label{exp22angolo}
  \end{center}
\end{figure}

In a second experiment, the dead load $P$ at the free end of the elastic rod of length $\bar{l}=0.45$ m (25 mm x 2 mm cross section, bending stiffness $B=2.70$ Nm$^2$) was imposed by filling two containers with water at a constant rate of 10 g/s, in order to obtain a slow and continuous increase in the applied load, which was measured with two miniaturized Leane XFTC301 (R.C. 500N) load cells.
The experimental results, expressed in terms of applied dimensionless load $p$ as a function of both the dimensionless deflection $u_2(\lbar)/\lbar$ (left) and amount of the rod inserted into the sliding sleeve, $l_{\textup{eq}}/\lbar$, (right) are reported in Fig. \ref{exp22angolo} for a dimensionless stiffness parameter $q=0.45$, together with the theoretically predicted behaviour.
Results reported as pink curves refer to a tilt angle $\alpha=9^\circ$, for which asymptotic self-restabilization occurs, whereas results reported as blue curves refer to a tilt angle $\alpha=20^\circ$, for which asymptotic self-restabilization does not occur. Experimental results are reported with continuous lines (from continuous measures) and dots or squares (from measures extracted from a video of the experiments through image post-processing,
for $\alpha=9^\circ$ and $\alpha=12^\circ$ respectively), together with the theoretical prediction (dashed line), showing a nice agreement between theory and experiments.

A movie showing the experiments considered in Fig. \ref{exp22angolo} is available as electronic supplementary material.

\section{Conclusion}

A structure which self-restabilizes, namely, capable of recovering its initial trivial path after a post-bifurcation deformation, would be useful for applications in soft robotics and
deformable mechanisms. One example of such a structure has been found in an asymptotic sense, so that a structural system has been shown to exhibit a deflection initially increasing from zero and later decreasing until vanishing. The asymptotic self-restabilization is determined by the effect of a configurational (or \lq Eshelby-like') force, which has been theoretically deduced from the principle of least action. The mechanical behaviour of the structure and the stability of the equilibrium paths have been theoretically solved and experimentally confirmed, so that the presented results may open new perspectives for technological applications.
The achievement of ({\sl non-asymptotic})  self-restabilization in the presence of linear elastic constraints still remains a challenge to be addressed.

\vspace*{5mm} \noindent
{\sl Acknowledgments } The authors gratefully acknowledge financial support from the ERC, Advanced Grant \lq Instabilities and nonlocal multiscale modelling of materials'
ERC-2013-AdG-340561-INSTABILITIES (2014-2019).
S.N. also acknowledges funding from the french `Agence Nationale de la Recherche' (ANR-14-CE07-0023-01).


\appendix*
\section{Lagrangian first variation}
%
%
Introducing the vector $\bm{w}=(x,y,\theta,\lin)$, collecting the functions describing the kinematics of the system, we consider the conditions for a state $\bm{w}$
to minimize the action ${\cal A}(\bm{w})$, under the boundary conditions (\ref{equa:BC-exit-point}) and fixed $x(s,t)$, $y(s,t)$, $\theta(s,t)$, and $\lin(t)$, at the two instants $t=t_{1}$ and $t=t_{2}$. Calculus of variations shows that a necessary condition for configuration $\bm{w}$ to be a solution is given by
\begin{equation}
{\cal A}'(\bm{w}) = \textbf{0},
\end{equation}
or equivalently
\begin{equation}
\left.  \frac{\d {\cal A}(\bm{w} + \epsilon \widetildeFV{\bm{w}})}{\d \epsilon}  \right|_{\epsilon=0}
= 0 \, , \quad \forall \, \widetildeFV{\bm{w}},
\end{equation}
where  $\widetildeFV{\bm{w}}=(\widetildeFV{x},\widetildeFV{y},\widetildeFV{\theta},\widetildeFV{\lin})$ is a variation of the configuration $\bm{w}$ such that
\begin{align}
\label{ctime}
\widetildeFV{\bm{w}}(s,t)=\Big(\widetildeFV{x}(s,t),\widetildeFV{y}(s,t),&\widetildeFV{\theta}(s,t),\widetildeFV{\lin}(t)\Big)=\textbf{0},\nonumber\\
& \mbox{for} \,\, t=t_1 \,\, \mbox{and} \,\, t=t_2 \,\, \forall s.
\end{align}
In addition to the conditions in time (\ref{ctime}), the boundary conditions (\ref{equa:BC-exit-point})
imply the following kinematical constraints on the variation $\widetildeFV{\bm{w}}$
\begin{equation}
\label{equa:BC-for-bar-terms}
\widetildeFV{x}(\lin) = - \widetildeFV{\lin} \: , \quad
\widetildeFV{y}(\lin) = 0 \: , \quad
\widetildeFV{\theta}(\lin) + \widetildeFV{\lin} \; \theta'(\lin)= 0.
\end{equation}
We perform integrations by parts with regard to time $t$ for the kinetic energy $\mathcal{T}$ and remark that the boundary terms vanish due to the condition (\ref{ctime}).
%
We then perform integrations by parts with regard to the arc-length
$s$ for the total potential energy $\mathcal{V}$ and finally obtain
\begin{align}
\left.  \frac{\d {\cal A}(\bm{w} + \epsilon \widetildeFV{\bm{w}})}{\d \epsilon}  \right|_{\epsilon=0}
 = & \int_{t_1}^{t_2}  a(\bm{w},\widetildeFV{\bm{w}}) \d t, \label{equa:Aprime}
\end{align}
with
 \begin{align}
a(\bm{w},\widetildeFV{\bm{w}}) &= \int_{-\hat{l}}^{\lbar} - \gamma ( \ddot{x} \,
\widetildeFV{x}+\ddot{y} \, \widetildeFV{y} \, ) \, \d s
+\frac{1}{2} B \, \widetildeFV{\lin} \, \theta'^2(\lin)
\nonumber\\& -P \cos \alpha \, \widetildeFV{x}(l) + P \sin \alpha \, \widetildeFV{y}(l)
\nonumber\\
&- k \, \left[x(-\hat{l}) +\hat{l}\, \right] \,  \widetildeFV{x}(-\hat{l})+\int_{\lin}^{\lbar} B \theta'' \, \widetildeFV{\theta} \,\d s \nonumber \\
&-\left[
B \theta' \, \widetildeFV{\theta} \, \right]_{\lin}^{\lbar}
+\int_{-\hat{l}}^{\lin}  N_x' \, \widetildeFV{x} \, \d s - \left[
N_x \, \widetildeFV{x} \right]_{-\hat{l}}^{\lin}
\nonumber\\
&-\int_{-\hat{l}}^{\lin}  N_x \, \sin \theta \, \widetildeFV{\theta} \, \d s  
%
+\int_{\lin}^{\lbar}  N_x' \, \widetildeFV{x} \, \d s - \left[ N_x
\, \widetildeFV{x} \right]_{\lin}^{\lbar}
\nonumber\\
&-\int_{\lin}^{\lbar}  N_x \, \sin \theta \, \widetildeFV{\theta} \, \d s
+\int_{-\hat{l}}^{\lin}  N_y' \, \widetildeFV{y} \, \d s - \left[
N_y \, \widetildeFV{y} \right]_{-\hat{l}}^{\lin}
\nonumber\\
&+\int_{-\hat{l}}^{\lin}  N_y \, \cos \theta \, \widetildeFV{\theta} \, \d s  
%
+\int_{\lin}^{\lbar}  N_y' \, \widetildeFV{y} \, \d s - \left[ N_y
\, \widetildeFV{y} \right]_{\lin}^{\lbar}
\nonumber\\
&+\int_{\lin}^{\lbar}  N_y
\, \cos \theta \, \widetildeFV{\theta} \, \d s,  \label{equa:a(w)0}
\end{align}
where the possibility of having a jump at the sliding sleeve exit point in the internal force $N_x(\lin)$ and $N_y(\lin)$
has been taken into account.
Note that for simplicity of presentation, the term $\theta'(\lin)$ in equation (\ref{equa:a(w)0}) and in the following, refers to the non-null value of the function at this point, that is just outside the sliding sleeve, $s=\lin^+$, so that we have $\theta'(\lin)=\theta'(\lin^+)$.
Using conditions (\ref{equa:BC-for-bar-terms}), a further manipulation of equation (\ref{equa:a(w)0}) leads to
\begin{align}
a(\bm{w},\widetildeFV{\bm{w}}) &=
\int_{\lin}^{\lbar}  (B \theta''-N_x \, \sin \theta+N_y \, \cos \theta)\, \widetildeFV{\theta} \, \d s
\nonumber\\
&+\int_{-\hat{l}}^{\lin}  \left[\left(N_x'- \gamma  \ddot{x} \right) \, \widetildeFV{x}+
\left(N_y' - \gamma  \ddot{y}\right)\, \widetildeFV{y}\right]\, \d s
\nonumber\\
&-\left[P\cos \alpha+N_x(\lbar)\right] \widetildeFV{x}(\lbar)
+ N_y(-\hat{l}) \, \widetildeFV{y}(-\hat{l})
 \nonumber\\
&+\int_{\lin}^{\lbar}  \left[\left(N_x'- \gamma  \ddot{x} \right) \, \widetildeFV{x}+
\left(N_y' - \gamma  \ddot{y}\right)\, \widetildeFV{y}\right] \, \d s
\nonumber\\
&
+ \left\{N_x(-\hat{l})- k \left[x(-\hat{l})+\hat{l}\right]\right\} \, \widetildeFV{x}(-\hat{l}) \nonumber \\
&
+\left\{\salto{0.4}{N_x(\lin) }
+ \frac{1}{2} B \, \theta'^2(\lin)
\right\}\, \widetildeFV{x}(\lin)
\nonumber\\
&+\int_{-\hat{l}}^{\lin}  (-N_x \, \sin \theta+N_y \, \cos \theta)\, \widetildeFV{\theta} \, \d s
\nonumber\\
&+\left[P\sin \alpha-N_y(\lbar)\right]  \widetildeFV{y}(\lbar)
-B \, \theta'(\lbar) \, \widetildeFV{\theta}(\lbar),
  \label{equa:a(w)}
\end{align}
which annihilation, for every variation $\widetildeFV{\bm{w}}=(\widetildeFV{x},\widetildeFV{y},\widetildeFV{\theta},\widetildeFV{\lin})$, yields
the equations (\ref{sys:dynamics}) governing the dynamics of the elastic system and the related boundary conditions (\ref{sys:dynamics-bcnd}).
In equation (\ref{equa:a(w)}) the symbol $\salto{0.4}{~ \cdot ~}$  denotes the jump of the relevant argument evaluated at a specific point, namely
\begin{equation}
\salto{0.4}{f(b)}=f(b^+)-f(b^-).
\end{equation}

In particular, requiring the first variation, equation (\ref{equa:Aprime}), to vanish for every rotation field $\widetildeFV{\theta}(s,t)$,
 and  for $t \in (t_1;t_2)$ yields, from equation (\ref{equa:a(w)}), the Elastica

\begin{align}
-N_x(s,t) \, \sin \theta(s,t)+ N_y(s,t) \, \cos \theta(s,t) =  0,\nonumber\\
 s\in [-\hat{l} ,\lin) \label{equa:balance-in-sleeve}\\
B \theta''(s,t)-N_x(s,t) \, \sin \theta(s,t)+N_y(s,t) \, \cos \theta(s,t) = & 0,\nonumber\\
s\in( \lin,\lbar \,],
\end{align}
together with the boundary condition
\begin{equation}
\theta'(\lbar,t) =0.
\label{equa16}
\end{equation}
On the other hand, imposing equation (\ref{equa:Aprime}) to be zero for every variation in the displacement fields $\widetildeFV{x}(s,t)$ and $\widetildeFV{y}(s,t)$,
and  for $t \in (t_1;t_2)$, yields the dynamic equations for the system along the $x$ and $y$ directions
\begin{align}
\label{equa-nx}
N_x'(s,t)= \gamma  \ddot{x}(s,t), \quad
N_y'(s,t) = \gamma  \ddot{y}(s,t), \nonumber\\ \qquad s\in[-\hat{l}, \lin) \cup( \lin,\lbar \, ],
\end{align}
as well as the translational equilibrium at specific points: at the loaded end 
\begin{equation}
N_x(\lbar,t)=-P \, \cos \alpha,\qquad
N_y(\lbar,t)=P \, \sin \alpha,
\label{equa18}
\end{equation}
at the end attached to the spring 
\begin{equation}
N_x(-\hat{l},t)=k \left[x(-\hat{l},t)+\hat{l}\right],
\label{bcd-nx}
\end{equation}
and at the sliding sleeve exit 
\begin{equation}
\salto{0.4}{N_x(\lin,t) }=- \frac{1}{2} B \, \theta'^2(\lin,t).
\label{config0}
\end{equation}
Equation (\ref{config0}) discloses that a non-null jump is present in the axial internal force at the exit of the sliding sleeve and that is provided
by the  presence of an outward configurational force $F_\textup{c}$ developed there,
\begin{equation}
\salto{0.4}{N_x(\lin,t) }=-F_\textup{c}.
\label{config2}
\end{equation}
Due to the linear elastic behaviour of the rod,
the bending moment is related to curvature through $M(s,t)=B \, \theta'(s,t)$, and the configurational force $F_\textup{c}$
can be rewritten as   equation (\ref{config}) similarly to
\cite{Bigoni-Bosi-DalCorso-Misseroni-2:2014}, \cite{inietto}, \cite{incapsulo}, \cite{Bosi-DalCorso-Misseroni-Bigoni:2014}.
We note that there is also a jump in the $y$ component of the internal force, $\salto{0.4}{N_y(\lin,t) } \neq 0$,  but that it is not prescribed and has to be evaluated once the solution in known.
The governing equation for the part of rod inserted in the sliding sleeve are here omitted.
We integrate equation (\ref{equa:null-curvature-in-sleeve}), using boundary conditions (\ref{equa:BC-exit-point}) and the inextensibility constraint, to obtain
\begin{equation}
y(s,t) = \theta(s,t) = 0 \, , \; \;   x(s,t)=s-\lin(t) \, , \; \; s \in [-\hat{l},\lin).
\end{equation}
From equation (\ref{equa:balance-in-sleeve}) we then have $N_y(s,t)=0$ for $s\in[-\hat{l}, \lin)$.
Next, we integrate equations (\ref{equa-nx}) and (\ref{bcd-nx}) to find
\begin{equation}
N_x(s,t) = -\gamma \, (s+\hat{l}\, ) \, \ddot{\lin}(t) - k \, \lin(t),\qquad s \in [-\hat{l},\lin)
\end{equation}
Finally using the jump condition (\ref{config}) we obtain the $x$ component of force just after the exit of the sliding sleeve
\begin{equation}
N_x(\lin^+,t)  = -F_\text{c} - k \, \lin - \gamma (\lin+\hat{l})\ddot{\lin}.
\label{equa24}
\end{equation}

\end{document}